\documentstyle[prl,aps,multicol,epsf]{revtex}
\twocolumn
\begin{document}
\draft
\title{
 Implementation of
universal quantum gates based on nonadiabatic geometric phases}
\author{Shi-Liang Zhu$^{1,3}$ and
Z. D. Wang$^{1,2,4}$\thanks{
To whom correspondence should be addressed.
Email address: zwang@hkucc.hku.hk}
}
\address{$^1$Department of Physics, University of Hong Kong, Pokfulam Road,
Hong Kong, China. \\
$^2$ Department of Material Science and Engineering,
University of Science and Technology of China, Hefei, China\\
$^3$ Department of Physics,
South China Normal University, Guangzhou, China\\
$^4$ Texas Center for Superconductivity, University of Houston, Houston,
Tx 77204\\
}
\address{\mbox{}}
\address{\parbox{14cm}{\rm \mbox{}\mbox{}
We propose an experimentally feasible  scheme to achieve quantum
computation based on nonadiabatic geometric phase shifts, in which
a cyclic  geometric phase is used to realize a set of universal
quantum gates. Physical implementation of this set of gates is
designed for Josephson junctions and for  NMR
systems. Interestingly, we find that the nonadiabatic phase shift
may  be independent of the operation time under appropriate
controllable conditions. A remarkable feature of the present
nonadiabatic geometric gates
is that there is no intrinsic
limitation on the operation time.
}}
\address{\mbox{}}
\address{\parbox{14cm}{\rm PACS numbers:
03.67.Lx, 03.65.Vf, 85.25.Cp}}

\maketitle

\newpage
\narrowtext

Quantum computers could efficiently solve 
 certain significant problems which are  intractable
for classical computers\cite{Shor}. The physical implementation of
quantum computation (QC) requires a series of accurately controllable
quantum gates. These gates may be  implemented experimentally  by
using  controlled dynamic or geometric operations.
 It is  remarkable that geometric operations based on adiabatic passages depend only on the
global feature of the path executed\cite{Berry}, and therefore
provides a possible fault-tolerant way to perform quantum gate
operations\cite{Zanardi,Jones,Duan,Falci}. Recently, several
schemes for adiabatic geometric QC were proposed
by using nuclear magnetic resonance(NMR)\cite{Jones}, trapped
ions\cite{Duan}, or superconducting nanocircuits\cite{Falci}. In
particular, an experimental realization of the conditional
adiabatic phase shift was reported\cite{Jones} with the NMR
technique. However, the distortions from the adiabatic
approximation were also seen in the NMR experiment. Moreover, the
adiabatic condition requires that the evolution time must be much
longer than the characteristic time $\tau_0$ of a qubit system;
while the evolution must be completed within the decoherence time.
This constraint leads to an intrinsic  limitation on the
operation time of quantum gates and seems to be a serious obstacle
to the physical implementation of some geometric QC
schemes, especially for those with solid state
systems\cite{Falci,Makhlin,Nakamura}, where the decoherence time
is very short.
 Even for the single qubit gate
operations based on the adiabatic Berry phase, they might be hard to
implement with the current experimental technique\cite{Zhu_JPC}.
Therefore, it is important to overcome
the operation time limitation in geometric QC set
by the adiabatic condition. A two-qubit nonadiabatic geometric
phase shift gate was proposed \cite{Wang} recently for NMR
systems. However, it is  still an interesting open problem to
design a general scheme to achieve the universal set of
nonadiabatic geometric gates which can be implemented in various
physical systems.

In this Letter, we propose a general scheme to achieve the
universal set of quantum gates based on cyclic nonadiabatic
geometric operations. The present scheme is experimentally
feasible, with the required experimental techniques not more
stringent than those for dynamical gate operations. Similar to the
case with adiabatic passages, the geometric gates based on
nonadiabatic cyclic operations also depend only on some global
features \cite{Aharonov}, which make them robust to certain
computational errors. 
Comparing with a nonadiabatic two-qubit gate proposed in Ref.[10], 
our scheme has at least two distinct advantages: (i)  both  a two-qubit gate and
two noncommutable single qubit gates in a realistic system
are designed based on a nonadiabatic geometric method,  
the latter being also highly nontrivial and useful in  
QC\cite{Duan}; (ii) the time-independent nonadiabatic phase 
shift may be realized. 
It is also remarkable that the set of gates designed are not only
all-geometric but also all-nonadiabatic, 
and the scheme  is applicable for several potential physical systems.



For universal QC, we need only to achieve two
kinds of noncommutable single-qubit gates, and one nontrivial
two-qubit gate \cite{Lloyd}. Before the design of geometric
quantum gates, we show first how to calculate the adiabatic and
non-adiabatic geometric phases.
For a spin-$1/2$ particle subject to an
arbitrary magnetic field ${\bf B}$, the nonadiabatic  cyclic
Aharonov-Anandan (AA) phase\cite{Aharonov} is just the solid angle
determined by the evolution curve in the projective Hilbert
space--a unit sphere $S^2$.
Any two-component 'spin' state
$|\psi\rangle=
[e^{-i\varphi/2}cos(\theta/2),\ e^{i\varphi/2}sin(\theta/2)]^T$
may be mapped into
a unit vector
${\bf n}=(sin\theta cos\varphi,
sin\theta sin\varphi,cos\theta)$
in the projective Hilbert space
via the relation
${\bf n}=\langle\psi|\stackrel{\rightarrow}{\sigma|}\psi\rangle$,
where $T$ represents the transposition of matrix.
By changing the magnetic field,
the AA phase is given by
$\gamma=-\frac{1}{2} \int_{C}
(1-cos\theta)d\varphi$,
where $C$ is along the actual evolution curve on
$S^2$, and is determined by
the equation: $\partial_t{\bf n}(t)=
-\mu{\bf B}(t)\times{\bf n}(t)/\hbar$.
This $\gamma$ phase recovers Berry phase in
adiabatic evolution\cite{Zhu}.

We then show how to achieve the universal set of quantum gates
based on nonadiabatic geometric AA phases. The single qubit
Hamiltonian $\hat{H}$ is chosen to go through a cyclic evolution
with period $\tau$ in the parameter space $\{ {\bf B} \}$. We
consider the process where a pair of orthogonal states
$|\psi_{\pm}\rangle$ can evolve cyclically. It is  necessary to
first decide the above cyclic evolution states. A phase difference
between $|\psi_{+}\rangle$ and $|\psi_{-}\rangle$ 
can be introduced by cyclically
changing $\hat{H}$. The phases acquired in this way would contain
both a geometric and a dynamical component. The dynamical phase
accumulated in the whole process can be removed by a simple 
method to be described  later\cite{Jones,Falci}, 
and thus only the geometric phase needs to be
considered at present. By taking into account the cyclic condition for
$|\psi_{\pm}\rangle$ and removing the dynamical phase, we have
the relation $U(\tau)|\psi_{\pm}\rangle =exp(\pm
i\gamma)|\psi_{\pm}\rangle$, where $U(\tau)$ is the evolution
operator. Here we have also used the result that $\gamma(-{\bf
n}(0))= -\gamma({\bf n}(0))$ at any time if the two initial states
correspond to $\pm {\bf n}(0)$\cite{Zhu}. We now write an
arbitrary  initial state  as
$|\psi_i\rangle=a_{+}|\psi_{+}\rangle+a_{-}|\psi_{-}\rangle$ with
$a_{\pm}=\langle \psi_{\pm}|\psi\rangle$, and express
the two cyclic initial states
as $|\psi_{+}\rangle = cos\frac{\chi}{2}|0\rangle+
sin\frac{\chi}{2}|1\rangle$ and
$|\psi_{-}\rangle = -sin\frac{\chi}{2}|0\rangle+
cos\frac{\chi}{2}|1\rangle$,
where $|0\rangle$ and $|1\rangle$
constitute the computational basis for
the qubit.
The final state at time $\tau$
is found to be $|\psi_f\rangle=U(\chi,\gamma)|\psi_i\rangle$, where
\begin{equation}
\label{single_U}
U=\left (
\begin{array}{ll}
e^{i\gamma}cos^2\frac{\chi}{2}+e^{-i\gamma}sin^2\frac{\chi}{2}
& isin\chi sin\gamma  \\
i sin\chi sin\gamma
&  e^{i\gamma}sin^2\frac{\chi}{2}+e^{-i\gamma}cos^2\frac{\chi}{2}
\end{array}
\right ).
\end{equation}

It is straightforward to verify that two operations
$U^1(\chi_1,\gamma_1)$ and $U^2(\chi_2,\gamma_2)$ are
noncommutable unless
$sin\gamma_1\sin\gamma_2sin(\chi_2-\chi_1)=0$. Since two kinds
of noncommutable operations constitute a universal set of
 single-bit gates, we achieve the
universal single-bit gates by choosing  $\chi_1\not=\chi_2$ $(mod\
2\pi)$ for any nontrivial phases $\gamma_1$ and $\gamma_2$. For
example, the phase-flip gate
$U_1=exp(-2i\gamma_1|1\rangle\langle1|)$ (up to an irrelevant
overall phase) is achieved at $\chi=0$; the gate
$U_2=exp(i\gamma_2\sigma_x)$ is obtained at $\chi=\pi/2$, which
produces a spin flip (NOT-operation) when $\gamma_2=\pi/2$ and an
equal-weight superposition of spin states when $\gamma_2=\pi/4$.

In terms of the computational
basis $\{|00\rangle$,$|01\rangle$,$|10\rangle$,
$|11\rangle \}$, the unitary operator to describe
the two-qubit gate is given by
$
U^{tq}=diag(U_{(\gamma^0,\chi^0)},U_{(\gamma^1,\chi^1)})
$
under the condition that the control qubit is far away from
the resonance condition for 
the operation of the target qubit. 
Here $\gamma^{\delta}$ $(\chi^\delta)$ represents the geometric
phase (the cyclic initial state) of the target qubit as long as the
state of the control qubit corresponds $\delta=0,1$ ($\delta$  denotes the
state of control qubit). Following Ref.\cite{Lloyd}, we find that
the unitary operator $U^{tq}$ 
is a nontrivial two-qubit gate
if and only if $\gamma^1 \not= \gamma^0$ or $\chi^1 \not= \chi^0$
$(mod\ 2\pi)$.
Therefore, all  elements of
QC may be achievable by using nonadiabatic cyclic
geometric operations.

  We now describe briefly how to remove the dynamical phase\cite{Jones,Falci}. 
 We let   $|\psi_{+/-}\rangle$
evolve along the time-reversal path of the first-period loop   during
the second period, namely, the same loop as before is covered backwards by
 $H_B=-\mu{\bf B}\cdot \stackrel{\rightarrow}{\sigma}/2 $. This process may  be simply realized  
by reversing the effective magnetic field with ${\bf B}(2\tau-t)=-{\bf B}(t)$
 on the same loop of the first period $[0, \tau)$.   
We thus have $H_B(2\tau-t)=-H_B(t)$ on the loop. 
As a result,
the geometric AA phases accumulated in the  
two periods will add and
the dynamical phase  will be cancelled.  This is because
the dynamical phase $\gamma_d^{(2)}$ for the second period has the same magnitude as that($\gamma_d^{(1)}$)
 for the first period but with the opposite sign, i.e.,  $\gamma_d^{(2)}=\int_\tau^{2\tau} dt \langle \psi|H_B(t)|\psi\rangle$ 
=$\int_0^\tau dt \langle \psi|H_B(2\tau-t)|\psi\rangle$= $- \int_0^\tau dt \langle \psi|H_B(t)|\psi\rangle$=
$- \gamma_d^{(1)}$, where the fact that $|\phi(t)\rangle=|\psi(2\tau-t)\rangle$ and $|\psi(t)\rangle$ represent
the same quantum state is used.

So far, we have proposed a general scheme to achieve nonadiabatic
geometric QC. It is important to further consider
implementing this scheme with real physical systems. Here, we
illustrate this implementation by two examples. The first one is an
NMR system\cite{Jones,Gershenfeld}, where the Hamiltonian for a
single qubit is given by
\begin{equation}
\label{rotated-field}
H=-\frac{1}{2}(\omega_0\sigma_x cos\omega t+\omega_0 \sigma_y
sin\omega t+\omega_1 \sigma_z),
\end{equation}
with $\omega_i=\mu B_i/h$. Equation (\ref{rotated-field}) can be solved
analytically\cite{Zhu}. In terms of explicit form of the 
solution ${\bf n}(\chi,\omega t)$ represented in Ref.\cite{Zhu},
it is found that the initial states $|\psi_{\pm}\rangle$
with $\chi=arctan[\omega_0/(\omega_1+\omega)]$ takes a cyclic
evolution with the period $\tau=2\pi/\omega$\cite{Wang},
 and the evolution
path is  the curve on a unit sphere swept by the unit vector
$\pm{\bf n}(\chi,\omega t)$. The corresponding geometric phase is
given by $\gamma_\pm =\pm\pi(1-cos\chi)$ for one
cycle.
Based on the noncommutable criterion
mentioned before, we may use any two processes with different
$\omega_0/(\omega_1+\omega)$ to achieve
two noncommutable qubit gates.
The advantage of the above
nonadiabatic gates is that there is in principle no limitation on
the magnitude of  $\omega$. It needs to be noted that,
 the gates $U_{1,2}$ may not be
practical by using the field ${\bf B}$ in Eq.(\ref{rotated-field})
since $\gamma=0$ $(\pi)$ as $\chi=0$ ($\pi/2$). This problem can
be solved by rotating  the field. It is seen that the parameter
$\chi$ for the initial cyclic state may be changed by  rotating
the symmetric axis for the field (\ref{rotated-field}). In the
rotated coordinates,  ${\bf B}^\prime={\bf R}(
\hat{y},\chi^\prime-\chi) {\bf B}$ (where ${\bf R}(
\hat{y},\theta)$ represents the rotation of  angle $\theta$
around  the $\hat{y}$-axis ) and ${\bf n}^\prime$ $(={\bf
R}(\hat{y},\chi^\prime-\chi){\bf n} (\chi,\omega t))$ because of
the spherical symmetry of the system. Thus $\chi$ may change to
any  required $\chi^\prime$ for  implementation of the quantum
gate, with the geometric phase being unchanged since the area
swept by ${\bf n}^\prime$ is the same as that by ${\bf n}$. For
example, if the magnetic field is ${\bf B}^\prime$ for
$\chi^\prime=\pi/2$, we may achieve the gate
$U_2=exp(i\gamma_2\sigma_x)$ with $\gamma_2=2\pi(1-cos\chi)$
(where the factor of 2 arises from the evolution of two cycles,
which is adopted to remove  the dynamic phase). A similar
method may be employed to achieve the two-qubit operation. The
spin-spin interaction in NMR is very well approximated
 by
$H_I=J\sigma_z^1\sigma_z^2/2$.  We may
use the initial state
$|\psi_{\pm}\rangle$ with
$\chi^{\delta}=arctan\{\omega_0/[\omega_1+(2\delta-1)J+\omega]\}$
   to achieve a nonadibatic  cyclic two-qubit
gate $U^{tq}$.
The state of control qubit is (almost) not
affected by any operation of the target qubit
if  $\omega_1^t$ of the target qubit is chosen to be
significantly different from $\omega_1^c$ of the control qubit
(i.e., $|\omega_1^t-\omega_1^c|>>J$).

\begin{figure}[tbp]
\label{fig1}
\vspace{-0.0cm}
\epsfxsize=8.5cm
\epsfbox{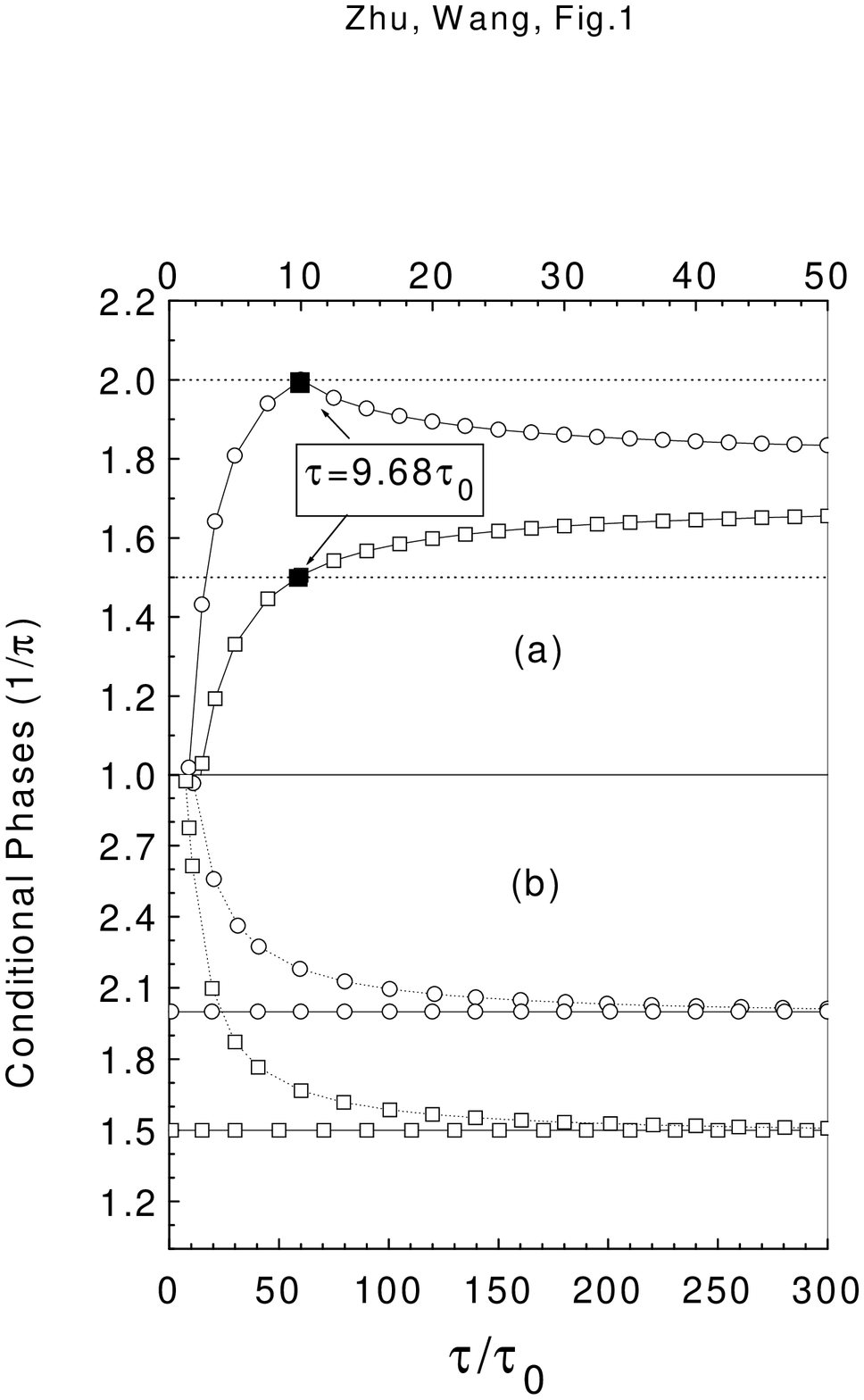}
\vspace{-1cm}
\caption{
Conditional geometric phases in an NMR system
for   (a) $\omega_1=0.8 J$ and (b) $\omega_1=J-\omega$ with $\omega_0=2\sqrt{15}J$.
The solid lines represent actual  phases for  cyclic evolutions,
while the dotted ones  are calculated under the adiabatic
approximation. Open circles  and squares denote 
$\gamma^0$ and $\gamma^1$,  respectively.
}
\end{figure}

As a typical case, we now show how to achieve
the conditional geometric phases for the two-qubit gate
$U^{tq}_{(2\pi,3\pi/2)}$.
$\gamma^\delta$ as a function of $\tau/\tau_0$
(with $\tau_0=2\pi/\omega_0$)
is plotted in Fig.1a and Fig.1b, where
$\omega_0$ is set to be $2\sqrt{15}J$.
First, we assume that $\omega_1=\alpha J$
with $0<\alpha<1$,
the phase shifts for $U^{tq}_{(2\pi,3\pi/2)}$
are calculated for $\tau=2\sqrt{15}/[(1-\alpha)J]$.
The curve $\gamma^\delta \sim \tau$ is plotted
 in Fig.1b for $\alpha=0.8$(the operation time
$\tau=5\sqrt{15}/(2 J)$).  Secondly, we choose $\omega_1=J-\omega$ if
$\omega_1$ is changeable, then the phase $\gamma^\delta$ would be
independent on the operation time. The time-independent
conditional phase shifts are clearly seen in Fig.1b,  namely, the
state always evolves along the same closed curve in the projected
Hilbert space for the chosen parameters. This also indicates the
geometric feature of nonadibatic AA phase. Normally,
 the state in the projected Hilbert space
is controlled by varying the effective field.

The second example is provided by the charge qubit using Josephson
junctions, which has been studied in Ref.\cite{Falci} with
adiabatic passages. The generalization to non-adiabatic quantum
computation is important for this kind of solid state qubits since
the decoherence time in these systems is typically short. The
single-qubit(Fig.2a) consists of a superconducting electron box formed by
an asymmetric SQUID with the Josephson coupling $E_{1}$ and
$E_{2}$, pierced by a magnetic flux $\Phi$ and subject to an
applied gate voltage $V_x=2en^e_x/C_x$ with $2en^e_x$ as the
offset charge and $C_x$ as the capacitance of the electron box. In
the charging regime (where $E_{1,2}<<E_{ch}$ with $E_{ch}$ as
charging energy) and at low temperatures, the system behaves 
as an
artificial spin-$1/2$ particle in a fictitious magnetic field
~\cite{Averin2}
\begin{equation}
\label{fictitious}
{\bf B}=\{ E_Jcos\alpha,-E_Jsin\alpha,E_{ch}(1-2n^e_x) \}
\end{equation}
where $E_J=\sqrt{E_{-}^2+4E_{1}E_{2}cos^2(\pi\Phi/\Phi_0)}$,
$tan\alpha=E_{-}tan(\pi\Phi/\Phi_0)/E_{+}$
with $E_{\pm}=E_1 \pm E_2$ and
$\Phi_0=h/2e$.
By changing $V_x$ and $\Phi$,
Eq.(\ref{fictitious}) draws a
curve in the parameter space
$\{ {\bf B} \}$. We here study a specific process described by
$\pi\Phi(t)=\Phi_0 atan[E_{+}tan(\omega t)/E_{-}]$,
$n_x^e(t)=[1-(E_Jctg\chi_0+\hbar\omega)/E_{ch}]/2$.
The fictitious field  is a
rotating field with a constant frequency(Fig.2b),
which guarantees that the angle
$\chi_0=arctan[E_J/(B_z(t)-\hbar\omega)]$
is time-independent.
We can find that the state denoted
by the vector
${\bf n}(\chi_0,-\omega t)$
undergoes a cyclic evolution with the period $\tau=2\pi/\omega$\cite{Wang}.
Therefore, we can obtain a  unitary operator
(\ref{single_U}) in the charge-qubit system, where $\chi=\chi_0$
and the AA phase $\gamma$ is the half solid angle
swept by the vector ${\bf n}(\chi_0,-\omega t)$.

The interaction between different charge qubits may be realized by
coupling  two asymmetric SQUIDS capacitively. If the coupling
capacitance $C_I$ is smaller than the others, the field on the
target qubit is given by Eq.(\ref{fictitious}), but the
z-component is replaced by
$B_z^\delta=E_{ch}(1-2n^e_{x})+E_I(n^e_{x,c}-\delta)$, where
$n^e_{x,c}$ represents the offset charge in the control qubit and
$E_I$ is the coupling energy\cite{Falci}. Obviously, the
$\gamma^\delta$ phase of the target qubit in the decoupled case is
different from $\gamma$ even though $(\Phi,n^e_{x})$ varies in the
same way. If the offset charge $n^e_{x,c}$ is time-independent in
the process , the state  ${\bf n}(\chi^{\delta},-\omega t)$ with
$\chi^\delta=arctan[E_J/(B_z^\delta-\hbar\omega)]$ still
undergoes a cyclic evolution. Thus the two-qubit operator $U^{tq}$
 may be obtained similar to the case in NMR. The
elimination of the adiabatic condition for two-bit geometric gates
is significant since the coherence time for two qubit gates is
typically much shorter than that for single qubit gates.

\begin{figure}[tbp]
\label{fig2}
\vspace{-0.6cm}
\epsfxsize=8.5cm
\epsfbox{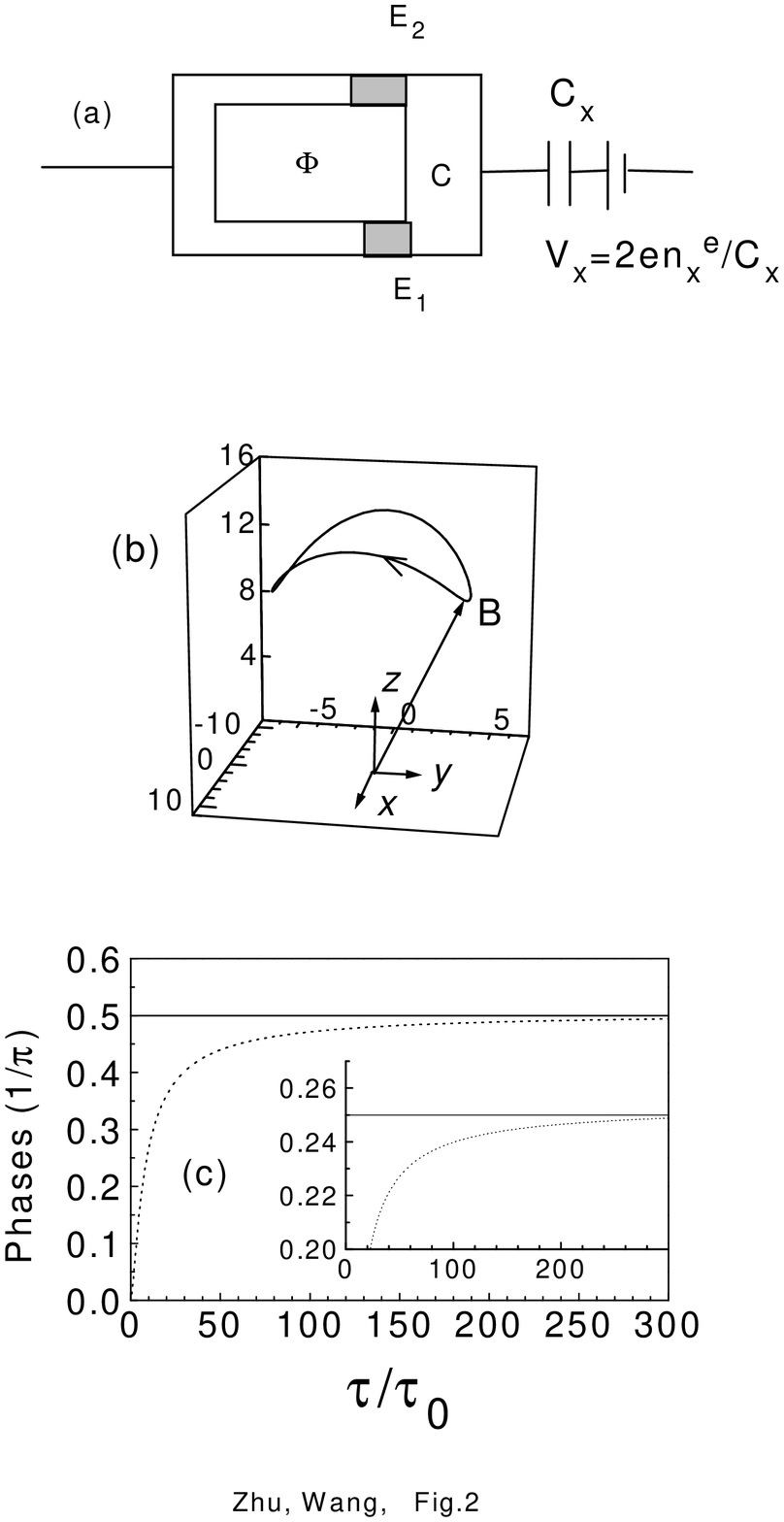}
\vspace{-0.3cm}
\caption{
(a) A schematic  Josephson charge qubit(Ref.[6]). 
(b) The fictitious magnetic field (in unit of $\mu ev$) in
 the designed cyclic process for nonadiabatic geometric gates.  
(c) Geometric phases to achieve a NOT gate
for $E_2=4E_1=6.25\mu ev$, $E_{ch}=5.0(E_1+E_2)$,
and $\chi_0=arccos(3/4)$. The inset shows the phase shift for the Hadamard
operation as $\chi_0=arccos(7/8)$. The solid lines represent the actual 
phases($\gamma$) for
cyclic evolutions,
while the dotted ones($\gamma_a$) are calculated under the adiabatic
approximation.
}
\end{figure}

 To see clearly the advantages of
nonadiabatic geometric qubit gates, we compare the operation time
between nonadiabatic gates and adiabatic gates, with the phase
shift $\pi/2$ $(\pi/4)$ corresponding to a NOT (Hadamard)
operation. The geometric phases for single-bit operations are
shown in Fig. 2c with certain parameters. It is seen from the main
panel of Fig. 2c that only if $\tau$ is longer than
 ~$70\tau_0$($=\hbar/E_J$) ,
 the  adiabatic phase $\gamma_a$
deviates from the actual nonadiabatic phase $\gamma$
within $10\%$ error. The similar results can be obtained
 for
other parameters(e.g., see the inset of Fig.2c for $\gamma=\pi/4$).
Note that the coherence time achieved by current technology is
$30\sim 40\tau_0$\cite{Nakamura} and thus the theoretical proposal
based on the adiabatic phase is
 not accurate enough to achieve this kind of  quantum gate experimentally.
In addition, Fig.2c  also clearly shows that
nonadiabatic phases can be independent of the operation time,
similar to the case in NMR systems addressed before.

Finally, we emphasize that the 
advantages of our scheme are: i.) the designed  quantum gates are  universal and
can handle arbitrary QC
without the intrinsic  limitation on operation time; ii.) the scheme is essentially 
all-geometric
and is robust to certain computational errors;
 iii.)  the physical implementation of the scheme
can now be realized by  realistic Josephson junctions and NMR systems.
Nevertheless, it seems to be a limitation
 of the method that the nonadiabatic phase
is more sensitive to fluctuation
in the trajectory area than that of the adiabatic phase.

This work was supported by the RGC grant of Hong Kong under
Nos. HKU7118/00P and HKU7114/02P,  the Ministry of Science and Technology of
China under No. G1999064602 and the URC fund of HKU.  ZDW thanks support in part from the
Texas Center for Superconductivity at the University of Houston.
We are very grateful to Drs. L. M. Duan and J. -W. Pan for their valuable
suggestions and remarks on the present work.
We also thank Prof. C. S. Ting for his critical reading of the manuscript.


\begin{references}


\bibitem{Shor} P. W. Shor, SIAM Rev. {\bf 41}, 303 (1999).

\bibitem{Berry} M. V. Berry, Proc. R. Soc. London {\bf A392}, 45
(1984).

\bibitem{Zanardi} P. Zanardi and M. Rasetti, Phys. Lett. A {\bf 264}, 94 (1999).

\bibitem{Jones} J. A. Jones, V. Vedral, A. Ekert, and G. Castagnoli,
Nature {\bf 403}, 869 (2000).


\bibitem{Duan} L. M. Duan, J. I. Cirac, and P. Zoller, Science
{\bf 292}, 1695 (2001).

\bibitem{Falci} G. Falci, R. Fazio, G. M. Palma,
J. Siewert, and V. Vedral, Nature {\bf 407}, 355 (2000).

\bibitem{Makhlin} Y. Makhlin, G. Sch\"{o}n, and A. Shnirman,
Rov. Mod. Phys. {\bf 73}, 357 (2001).
\bibitem{Nakamura} Y. Nakamura, Yu. A. Pashkin, and J. S. Tsai,
Nature {\bf 398}, 786 (1999).


\bibitem{Zhu_JPC} S. L. Zhu and Z. D. Wang, Physica C {\bf 364}, 213 (2001).



\bibitem{Wang} X. B. Wang and M. Keiji, Phys. Rev. Lett.
{\bf 87}, 097901 (2001); {\bf 88}, 179901(E) (2002).


\bibitem{Aharonov} Y. Aharonov and J. Anandan, Phys. Rev.
Lett. {\bf 58}, 1593 (1987).


\bibitem{Lloyd} S. Lloyd, Phys. Rev. Lett. {\bf 75}, 346 (1995).

\bibitem{Zhu} S. L. Zhu and Z. D. Wang, Phys. Rev. Lett. {\bf 85}, 1076
(2000).







\bibitem{Gershenfeld} N. Gershenfeld and I. L. Chuang,
Science {\bf 275}, 350 (1997).



\bibitem{Averin2} D. V. Averin and K. K. Likharev,
in B. L. Altshuler, P. A. Lee, and R. A. Webb,
Mesoscopic Phenomena in Solids (Elsevier, New York, 1991), p213.





\end{references}
\end{document}